# VEPerform: a web resource for evaluating the performance of variant effect predictors

Cindy Zhang[1,2,3], Frederick P. Roth[1,2,3]

[1] Department of Computational and Systems Biology, University of Pittsburgh School of Medicine, Pittsburgh, PA, USA
[2] Donnelly Centre and Department of Molecular Genetics, University of Toronto, Toronto, ON, Canada
[3] Lunenfeld-Tanenbaum Research Institute, Sinai Health, Toronto, ON, Canada

## Abstract

### Summary

Computational variant effect predictors (VEPs) are providing increasingly strong evidence to classify the pathogenicity of missense variants. Precision vs. recall analysis is useful in evaluating VEP performance, especially when adjusted for imbalanced test sets. Here, we describe VEPerform, a web-based tool for evaluating the performance of VEPs at the gene level using balanced precision vs. recall curve (BPRC) analysis.

### Availability

VEPerform service: cdyzh.shinyapps.io/veperform/.
VEPerform source code: https://github.com/cdyzh/VEPerform.

## Introduction

Understanding the impact of genetic variants is critical in biology and, in a clinical setting, variants of uncertain significance (VUS) are limiting the promise of precision medicine. The majority of currently-reported VUS are missense and, conversely, the majority of missense variants are classified as VUS[1,2]. Fortunately, computational variant effect predictors (VEPs) are providing ever-stronger evidence in classifying the pathogenicity of missense variants, as supported by VEP evaluations that aggregate performance across genes and diseases[3–6]. However, there is a need for a standardized platform to assess VEP performance at the individual gene level. This is useful not only to choose an appropriate VEP for each gene, but also to begin to evaluate the strength of evidence they can provide. VEPerform is a web-based app that allows users to evaluate the performance of different missense VEPs for individual genes.

## Methods for evaluating performance

VEPerform assesses performance of each VEP by evaluating, across the range of potential score thresholds for that VEP, the tradeoff between precision (fraction of variants below a given threshold that are truly pathogenic) vs recall (fraction of known pathogenic variants that are identified using that threshold). Because precision depends quantitatively on balance (e.g., proportion of pathogenic variants) within the reference set, VEPerfom transforms precision values into balanced precision values (the precision that we would expect to have observed if the test set had been balanced to contain exactly 50% pathogenic variants)[7]. From the resulting balanced precision vs recall curves (BPRCs), two key single-point performance measures can be extracted: 1) The recall achieved at a stringent (90%) balanced precision threshold (R90BP); 2) the area under the BPRC curve (AUBPRC).

## Basic Mode

In Basic mode, users select a specific gene, then choose one or more of the default predictors (currently VARITY, REVEL, and AlphaMissense). A key element of this analysis is of course the reference set of variants used in the evaluation. For each gene, VEPerform provides a default reference set of ClinVar curated variants. These reference sets, as of July 2024, collectively included 175,165 variants annotated as pathogenic, likely-pathogenic, benign, or likely-benign, as well as 76,197 variants with conflicting interpretations and 1,193,162 VUS.

Because individual users may have knowledge about which reference variants are reliable, were not used in training the VEPs, and are most representative of variants they seek to classify (e.g., variants causing a specific disease phenotype), VEPerform allows users to filter out variants in two ways: First, users can select to filter out common variants (defined by having minor allele frequency > 0.005 in gnomAD, v3). Second, after users initiate the analysis by clicking "Make Precision vs Recall Plot", a pop-up window opens in which users can see each variant's HGVSp descriptor and ClinVar review status and remove any problematic variants.

Users may download both the BPRC plot and all input and metadata used in the analysis into a PDF report. Users can also download (as a CSV file) all variants that were used for BPRC analysis, as well as the broader list of variants (e.g., including previously filtered, VUS, and conflicting variants). This download option enables users to bring their own domain knowledge to bear in Advanced mode, as described next.

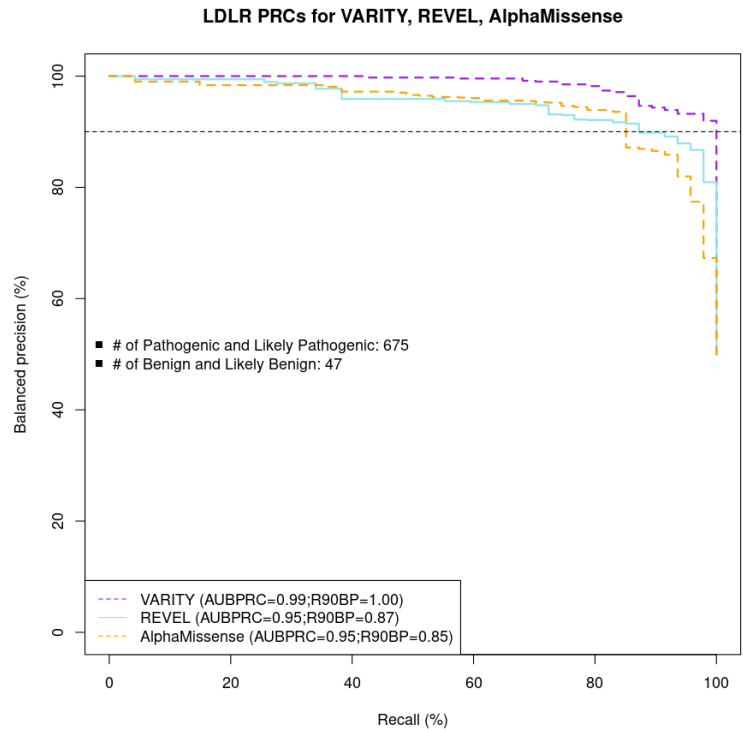

Figure 1: the web interface of VEPerform Basic Mode

## Advanced Mode

The Advanced mode of VEPerform provides flexibility for users, in that they can further customize the list of variants, provide variant annotations from their own sources, and provide (or request that VEPerform retrieve) scores from additional VEPs. To load customized input data into VEPerform, users may either edit the variant information file downloaded using Basic Mode, or generate their own variant information file. In either case, Advanced Mode requires that the user upload a variant information file including at least two columns: gene name and variant pathogenicity annotation (either "P/LP" for variants to be treated as pathogenic in the evaluation or "B/LB" for variants to be treated as benign).

The file may also include any number of additional columns for the scores from additional VEPs (or quantitative experimental measures of variant function). Where rows for new variants beyond the default variant set have been added, or the user wishes to evaluate VEPs outside of the default VEP set, users should fill in the missing VEP scores, providing a new column for each new VEP. Alternatively, if the user instead provides the four columns needed to describe each variant's genomic coordinates, VEPerform will use these coordinates to query OpenCRAVAT's API in real-time and retrieve scores from any VEP selected from a menu of VEPs offered by OpenCRAVAT.

Because VEPerform's common variant filter relies on gnomAD allele frequencies, users also can either choose to provide the allele frequency or, if genomic coordinates have been provided, fetch this data from OpenCRAVAT. Finally, Advanced Mode allows all download options that were described above for Basic Mode.

## Caveats

An important caveat for VEPerform users is the potential for inflated performance estimates due to circularity, i.e., where a VEP was both trained using and overfit to variants in VEPerform's reference set. Having initially identified VEPs as high-performing based on either the default or a user-selected reference variant set, users should therefore (where possible) also evaluate performance without reference variants used in training. This is less of an issue for VEPs that have not been trained on pathogenicity annotations such as EVE[8], ESM1-b[9], or ESM1-v[10], or AlphaMissense[11], or for supervised VEPs such as VARITY[7] that have been shown to avoid overfitting[6]. Because some VEPs are substantially better at distinguishing pathogenic variants from benign variants that are common than those that are rare[7,12], we also recommend that users evaluate performance after excluding common variants.

## Conclusion

VEPerform enables users to carry out gene-focused evaluation of variant effect predictors. By leveraging a combination of pre-stored data and custom user-uploaded variant processing, enables biomedical researchers and clinical geneticists to incorporate their domain knowledge to make informed decisions about VEPs most appropriate for their application. Where all VEPs perform poorly, VEPerform can help prioritize genes for functional testing.

## Acknowledgements

This work was supported by the National Human Genome Research Institute of the National Institutes of Health (NHGRI) Center of Excellence in Genomic Science Initiative (F.P.R.; RM1HG010461). F.P.R was also supported by the NHGRI Impact of Genomic Variation on Function Initiative (F.P.R.; UM1HG011989), by R01HL164675 (F.P.R., D.M.R.), and by a Canadian Institutes of Health Research Foundation Grant (F.P.R.; FDN-159926).